\documentclass[conference]{IEEEtran}

\IEEEoverridecommandlockouts
\usepackage{cite}
\usepackage{amsmath,amssymb,amsfonts}
\usepackage{algorithmic}
\usepackage{graphicx}
\usepackage{textcomp}
\usepackage{xcolor}
\usepackage{soul}
\usepackage{breqn}
\usepackage{flushend}

\usepackage{tabularx}
\usepackage[subrefformat=parens]{subcaption}
\usepackage{comment}
\usepackage{url}

\def\BibTeX{{\rm B\kern-.05em{\sc i\kern-.025em b}\kern-.08em
    T\kern-.1667em\lower.7ex\hbox{E}\kern-.125emX}}

\title{JABBERWOCK: A Tool for WebAssembly Dataset Generation and Its Application to Malicious Website Detection}

\makeatletter
\newcommand{\linebreakand}{%
  \end{@IEEEauthorhalign}
  \hfill\mbox{}\par
  \mbox{}\hfill\begin{@IEEEauthorhalign}
}
\makeatother

\author{\IEEEauthorblockN{Chika Komiya}
\IEEEauthorblockA{Osaka University, Japan \\
c-komiya@ist.osaka-u.ac.jp}
\and
\IEEEauthorblockN{Naoto Yanai}
\IEEEauthorblockA{Osaka University, Japan \\
yanai@ist.osaka-u.ac.jp}
\and
\IEEEauthorblockN{Kyosuke Yamashita}
\IEEEauthorblockA{Osaka University, Japan \\
yamashita@ist.osaka-u.ac.jp} 
\linebreakand
\IEEEauthorblockN{Shingo Okamura}
\IEEEauthorblockA{National Institute of Technology, Nara College, Japan \\
okamura@info.nara-k.ac.jp}
}

\begin{document}

\maketitle

\begin{abstract}
Machine learning is often used for malicious website detection, but an approach incorporating WebAssembly as a feature has not been explored due to a limited number of samples, to the best of our knowledge.
In this paper, we propose \textit{JABBERWOCK (JAvascript-Based Binary EncodeR by WebAssembly Optimization paCKer)}, a tool to generate WebAssembly datasets in a pseudo fashion via JavaScript. 
Loosely speaking, JABBERWOCK automatically gathers JavaScript code in the real world, convert them into WebAssembly, and then outputs vectors of the WebAssembly as samples for malicious website detection. 
We also conduct experimental evaluations of JABBERWOCK in terms of the processing time for dataset generation, comparison of the generated samples with actual WebAssembly samples gathered from the Internet, and an application for malicious website detection. 
Regarding the processing time, we show that JABBERWOCK can construct a dataset in 4.5 seconds per sample for any number of samples. 
Next, comparing 10,000 samples output by JABBERWOCK with 168 gathered WebAssembly samples, we believe that the generated samples by JABBERWOCK are similar to those in the real world. 
We then show that JABBERWOCK can provide malicious website detection with 99\% F1-score because JABBERWOCK makes a gap between benign and malicious samples as the reason for the above high score. 
We also confirm that JABBERWOCK can be combined with an existing malicious website detection tool to improve F1-scores. 
JABBERWOCK is publicly available via GitHub (\url{https://github.com/c-chocolate/Jabberwock}). 
\end{abstract}


\section{Introduction} 
\label{sec:introduction}
%
Malicious websites have caused various cyber crimes, such as phishing and cryptojacking. 
While malicious websites can be detected by blacklisting websites conventionally, machine learning has been introduced in recent years due to the high occurrence of new malicious websites. 
It can detect even new kinds of malicious websites by training a model with their features. 

Although various features have been introduced until now, we focus on \textit{WebAssembly} in this paper. 
Loosely speaking, WebAssembly is a language designed for running on a virtual machine and has been implemented on four major browsers, i.e., Firefox, Chrome, Safari, and Microsoft Edge, nowadays.
According to Battagline~\cite{battagline2021art}, the execution of WebAssembly codes is at most eight times faster than that of JavaScript codes, and there are websites that potentially accept WebAssembly codes. 
Websites with WebAssembly will increase in the future~\cite{Musch}. 
Thus, it is considered that the use of WebAssembly as a feature is desirable for malicious website detection. 

In this paper, we answer the following question about malicious website detection: \textit{is malicious website detection based on WebAssembly possible?}
The above question is non-trivial due to the following problems. 
First, to the best of our knowledge, there is no publicly available dataset of WebAssembly. 
We note that the design of a WebAssembly dataset is challenging because the number of websites with WebAssembly is limited at present. 
In general, malicious website detection needs to collect and learn samples of WebAssembly for benign and malicious websites. 
Second, even if they are collected, we will need to identify whether malicious website detection based on WebAssembly can provide high accuracy through experimental evaluations. 

To tackle the above problems where no WebAssembly dataset is available, we first propose a novel tool, \textit{JABBERWOCK (JAvascript-Based Binary EncodeR by WebAssembly Optimization paCKer)}, to generate WebAssembly datasets for malicious website detection. 
In a nutshell, JABBERWOCK collects JavaScript codes from existing websites and then converts them to WebAssembly. 
Afterward, JABBERWOCK returns a dataset by vectoring WebAssembly with natural language processing. 
Hence, JABBERWOCK generates WebAssembly datasets for malicious website detection in a pseudo fashion because it can collect and convert JavaScript codes from benign and malicious websites. 
JABBERWOCK is publicly available via GitHub (\url{https://github.com/c-chocolate/Jabberwock}). 

We then conduct experiments to evaluate WebAssembly datasets generated by JABBERWOCK and their resultant malicious website detection. 
Specifically, we gathered 10,000 samples through the existing crawling websites, and then show that JABBERWOCK can generate a dataset in 4.5 seconds per sample. 
We also demonstrate that samples in the generated dataset by JABBERWOCK are statistically similar to WebAssembly codes in the real world. 
Next, when we evaluate the malicious website detection with the dataset generated by JABBERWOCK, malicious websites can be detected with 99\% F1-score. 
We also found that JABBERWOCK makes a gap between benign and malicious samples as the reason for such a high score. 
We confirm that an existing malicious website detection tool~\cite{MADMAX} can be improved with the WebAssembly samples generated by JABBERWOCK in comparison with the latest work of malicious website detection~\cite{janaka2023madonna}. 
Our promising results show that malicious website detection is possible. 

To sum up, we make the following contributions in this paper: 
\begin{itemize}
    \item We propose JABBERWOCK, a publicly available tool to generate WebAssembly datasets from JavaScript code for malicious website detection. 

    \item JABBERWOCK can generate a dataset in 4.5 seconds per sample. 

    \item The dataset generated by JABBERWOCK is similar to WebAssembly codes in the real world.

    \item Malicious websites can be detected with 99\% F1-score according to experimental evaluations with JABBERWOCK. 

    \item The reason for the above high score is that JABBERWOCK makes a gap between malicious and benign samples. 

    \item An existing malicious website detection tool can be improved with the WebAssembly samples generated with JABBERWOCK. 

\end{itemize}

This paper is the full version of our previous work published~\cite{komiya2023jabberwock} which will be presented at DCDS~2023. 
In the previous work, we designed the dataset generation tool, JABBERWOCK, and then showed that WebAssembly datasets generated by JABBERWOCK are similar to WebAssembly codes in the real world. 
In this paper, we further evaluate malicious website detection based on the WebAssembly datasets and then demonstrate that malicious websites can be detected with a high F1-score. 
We also combine JABBERWOCK with an existing tool for malicious website detection and then show that F1-score can be improved. 
They were future works mentioned in the previous work. 
We believe that these future works were solved in this paper. 

\section{Backgrounds} \label{sec:backdgrounds}

This section describes  the backgrounds of WebAssembly and malicious website detection. 
We also describe related works. 

\subsection{WebAssembly} 
\label{sec:webassembly}

WebAssembly is a virtual instruction set architecture for stack machines. 
Generally, instruction set architecture is a binary format that aims to run on some specific machines.
Namely, WebAssembly runs on virtual machines, and thus it runs on variants of computer hardware and devices. 
Further, when deploying WebAssembly as part of Web applications, the download time will be short due to the small size of a binary file. 

To use WebAssembly on websites, we should put WebAeemlby Text (WAT) on a web server, where WAT is a WebAssembly file compiled from a pseudo-Assembly language. 
We can use WebAssembly by calling it via a JavaScript code. 

We remark that WebAssembly has variants of advantages. 
It does not depend on platforms, and has been implemented on four major browsers~\cite{Musch}.
Further, the execution of WebAssembly codes is at most eight times faster than that of JavaScript codes~\cite{battagline2021art}.
There are libraries that convert other languages to WebAssembly, such as wasm-pack\footnote{\url{https://github.com/rustwasm/wasm-pack}} for Rust, and Emscripten\footnote{\url{htps://emscripten.org}} for C/C+.
Thus, we believe that WebAssembly will be widely used on web pages. 


\subsection{Malicious Website Detection}

There are two main approaches to detecting malicious websites, i.e., knowledge-based and machine learning-based methods~\cite{Zhauniarovich2018survey}. 
The knowledge-based approach is heuristics and distinguishes benign and malicious websites by an expert's knowledge. 
Due to the continuous development of new malicious websites, it often fails. 
On the other hand, the machine learning-based approach infers unseen websites as benign or malicious~\cite{PALANIAPPAN2020654} through training a model with labeled websites. 
Hence, we focus on malicious website detection based on machine learning. 

Informally, malicious website detection based on machine learning involves training a model to learn the features of websites and their labels, which represent whether the websites are benign or malicious. 
During the inference phase, the model takes features of a target website as input and then infers its label as either benign or malicious.


\subsection{Related Work} 

\subsubsection{Design of WebAssembly Datasets}

The closest works are Wobfuscator~\cite{Wobfuscator} and the empirical study by Hilbig et al.~\cite{hilbig2021empirical}.  
Wobfuscator is a tool to obfuscate JavaScript malwares to WebAssembly, and is motivated by malware analysis. 
On the other hand, the empirical study by Hilbig~\cite{hilbig2021empirical} collected WebAssembly from live websites. 
However, the above datasets do not contain correctly labeled samples of WebAssembly collected from live websites as benign or malicious, which are important for machine learning-based malicious website detection.  
Many existing works are based on empirical analysis for vulnerabilities~\cite {lehmann2020everthing} and malwares~\cite{MINOS}. 
They are different from our work. 
The recent work~\cite{lehmann2022finding} aims to analyze types of functions and hence there is no publicly available dataset for malicious website detection.
Our goal is to design a dataset for malicious website detection.

\subsubsection{Features of Malicious Website Detection}

In malicious website detection, models that use only domain names have been widely used, although it requires high computational power~\cite{yu2018character,berman2019dga,yang2020detecting,khalil2018adomainisonlyasgood}. 
In addition to domain names which are just text data, other information can be used such as DNS information~\cite{10.1109/DSN.2015.35,10.1145/2960409,9110462,10.1145/2897845.2897877,242062} and web contents~\cite{7473534,10.1145/3029806.3029819,10.1145/3372297.3417233,ariyadasa2022combining,senanayake2021android}.
In this work, we suggest utilizing WebAssembly as a new feature for malicious website detection. 
We primarily focus only on WebAssembly. 
We also combine it with other existing features.

\subsubsection{Further Related Work}
Recently, malicious website detections have been extended to browser-based applications~\cite{tang2021surveyofmachinelearning}. 
For instance, several plug-ins~\cite{armano2016real-time,marchal2017off-the-hook} have been developed for phishing website detection. 
However, they utilize white-/blacklists of domains as a part of detection. 
The first work for a browser-based application utilizing only machine learning is MADMAX~\cite{MADMAX}, including feature selection. 
In recent years, MADDONA~\cite{janaka2023madonna} was proposed as an extension of MADMAD to optimize feature selection and neural network architectures further in accordance with the latest feature engineering~\cite{alhogail2022improveddetection}. 
We combine our dataset with MADMAX and then show that its resultant performance outperforms MADDONA. 
We leave it as an open problem to design a browser-based application with WebAssembly.



\section{Design of JABBERWOCK} 
\label{sec:design_of_jabeerwock}

Now we propose JABBERWOCK. 
We first overview JABBERWOCK, and then describe it in detail.
We finally explain the problem setting. 

\subsection{Overview}\label{sec:overview}

JABBERWOCK generates a dataset for machine learning-based  malicious website detection which uses WebAssembly as a feature. 
As mentioned in Section~\ref{sec:introduction}, the number of samples of WebAssembly is limited at present.
Therefore, we construct a pseudo dataset by converting JavaScript code, which is a major language for websites, into WebAssembly and then vectorizing WebAssembly as a sample for the dataset.
In doing so, JABBERWOCK takes a list of URLs as input and returns vectors of WebAssembly as output. 

\begin{figure}[t]
  \centering
  \includegraphics[width=1\linewidth]{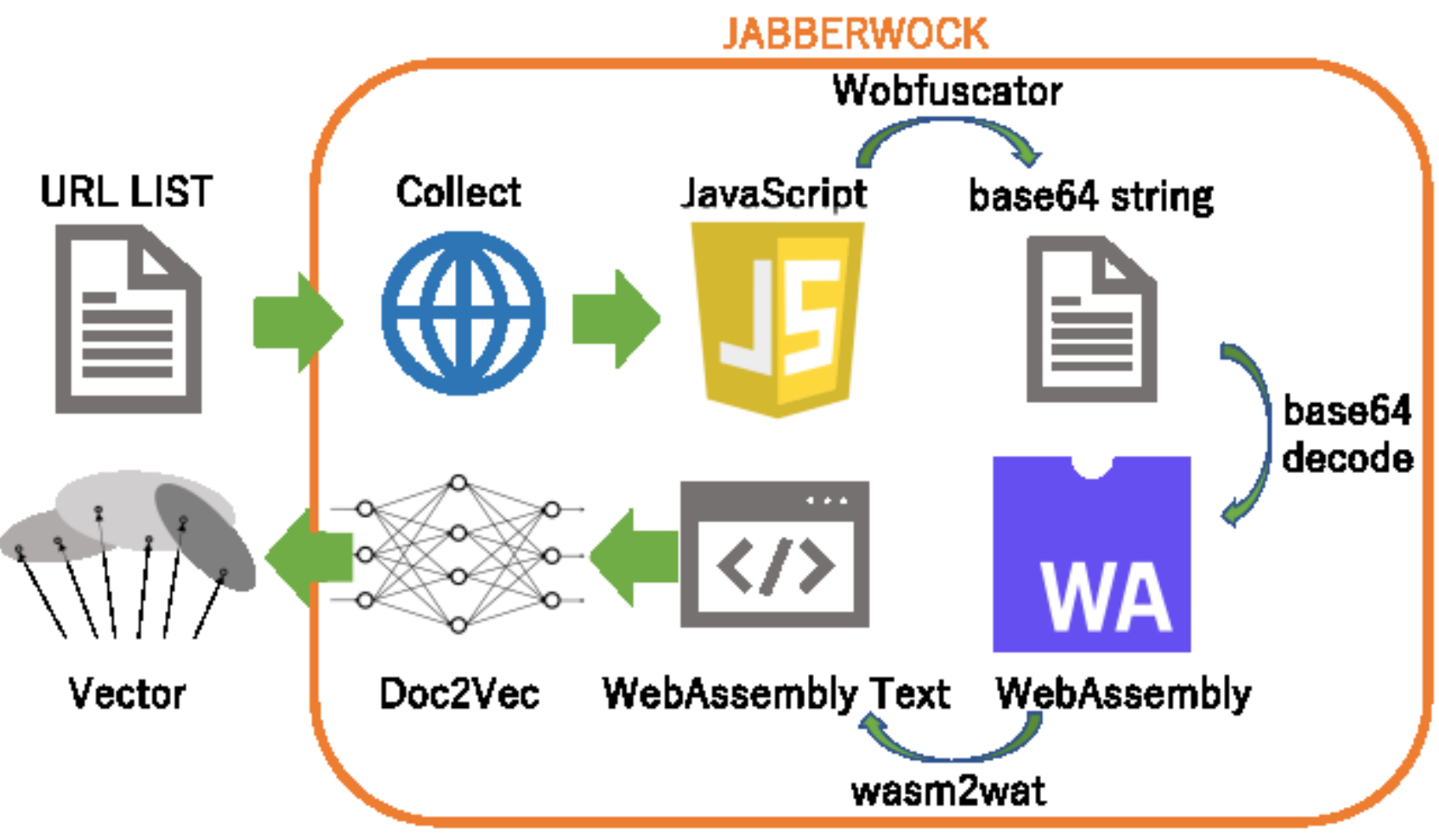}
  \caption{Overview of JABBERWOCK}
  \label{fig:overview_of_JABBERWOCK}
\end{figure}

Specifically, JABBERWOCK consists of the following three processes: 
(1) collect JavaScript codes from a given URL list; 
(2) convert them into WebAssembly; 
(3) vectorize WebAssembly through training a natural language processing model.
Fig.~\ref{fig:overview_of_JABBERWOCK} shows the overview of JABBERWOCK. 

JABBERWOCK accepts both labeled and unlabeled URL lists as inputs. 
Namely, when an URL is labeled (either benign or malicious), it outputs a vector with the same label. 
We remark that an URL list can consist of both benign and malicious websites. 
We follow the existing work~\cite{MADMAX} to label the samples: for instance, samples generated from URLs in main popularity ranking websites are benign and those from URLs in existing blacklists are malicious. 




\subsection{Detailed Description}

\subsubsection{Collection of JavaScript}

JABBERWOCK first collects JavaScript codes from a given URL list, which can be obtained from a popularity ranking such as Tranco~\cite{tranco}.
In the collection of JavaScript, JABBERWOCK finds pages of websites from the given URL lists, including benign and malicious websites, and then extracts a single file of JavaScript code for each page. 
Whereas it can extract multiple files of JavaScript code, the current specification extracts only a single file to suppress data bias. 
We implemented this function with \texttt{requests} library version 2.27.1 in Python. 

\subsubsection{Conversion into WebAssembly} 


We use Wobfuscator~\cite{Wobfuscator} to convert the collected JavaScript codes into WebAssembly. 
It is a tool to partially translate JavaScript code, which is potentially expressed by WebAssembly, into Base64 strings.\footnote{The original motivation of Wobfuscator is to convert into WebAssembly, but the current tool outputs Base64 strings.}
These strings can be decoded to \texttt{.wasm} files, and then are converted into WAT files with the \texttt{wasm2wat} package\footnote{\url{https://www.npmjs.com/package/wasm2wat}}. 
This will convert them to a format suitable for natural language processing in the next section.


\subsubsection{Vectorization with Natural Language Processing}

We use Doc2Vec\footnote{\url{https://radimrehurek.com/gensim/models/doc2vec.html}} as a natural language processing model to vectorize WebAssembly. 
Doc2Vec is a model to vectorize any given text. 
It is trained with the converted files of WAT by splitting into each line, and returns vectors as samples of a dataset. 
It can also be extended into any paragraph vector model.

\subsection{Problem Setting} 

The issues discussed in this paper are twofold.
We identify how vectors in a dataset output by JABBERWOCK are close to those of actual WebAssembly in the real world. 
Hereafter, we denote by J2W the data obtained by JABBERWOCK with JavaScript and by OWA the (unlabeled) original samples of WebAssembly. 
We also denote by J2W-based model and by OWA-based model as models trained with J2W and OWA, respectively. 
Our goal is to confirm whether a J2W-based model can infer websites similarly to an OWA-based model. 
Furthermore, we will check whether malicious website detection is actually possible with the JABBERWOCK dataset.


\section{Experiments} 

In this section, we conduct experiments with JABBERWOCK.
First, we describe the purpose and experimental setting, followed by an evaluation of JABBERWOCK in terms of processing time and difference from samples in the real world.
Next, we show the accuracy of the detection of malicious websites as an application of this method.

\subsection{Purpose of Experiments}\label{sec:purpose_of_experiments}

The purpose of the experiments is to evaluate JABBERWOCK in terms of the following three standpoints: 
(1) how long is the processing time to generate a dataset?; 
(2) is the distribution of the generated dataset close to the distribution of a real dataset?; 
(3) are datasets generated by JABBERWOCK potentially effective for malicious website detection?
We conduct three experiments based on the standpoints above.

First, the processing time is measured with respect to the number of data samples in terms of the collection of JavaScript codes, the conversion into WebAssembly, and the vectorization with natural language processing. 
Hence, we evaluate the processing time required for the dataset generation. 



\begin{figure}[tb]
 \centering
  \includegraphics[width=1\linewidth]{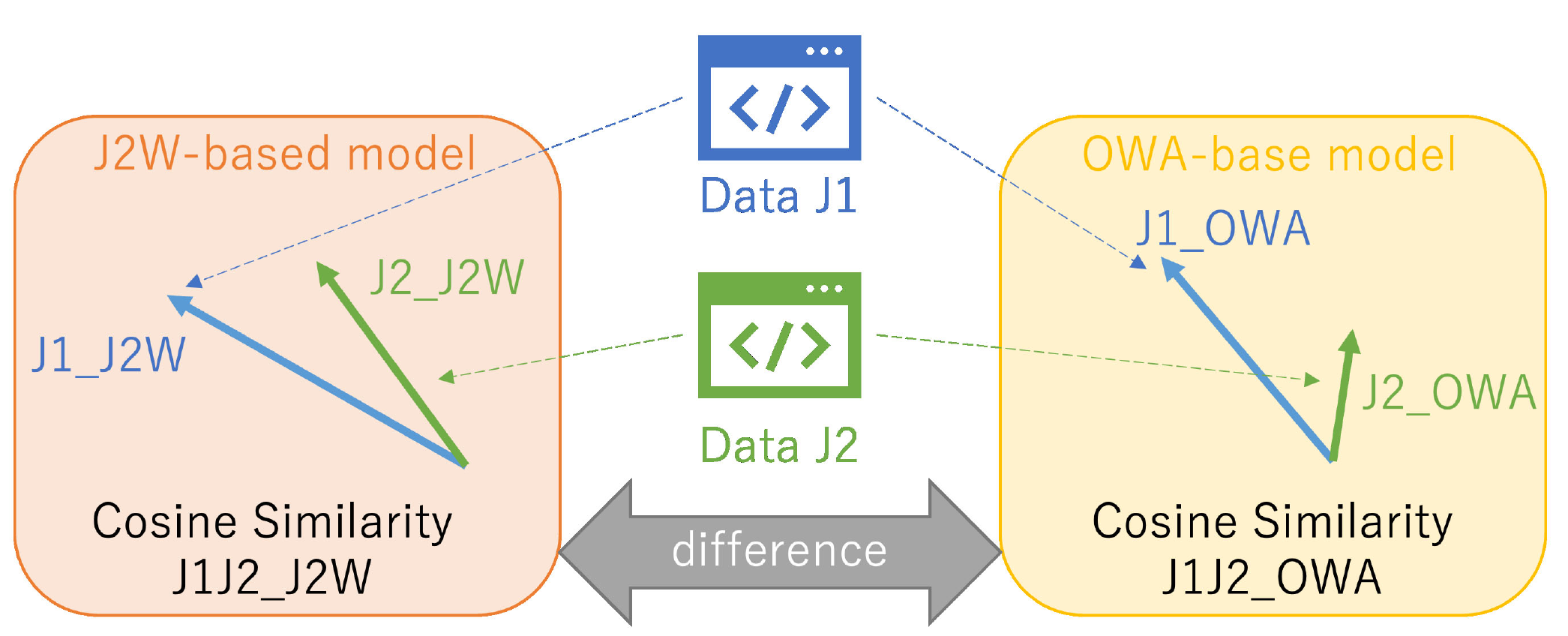}
  \caption{Intuition of evaluation metric for the distribution of datasets: 
   The variables, $\mathrm{J}_1, \cdots, \mathrm{J}_n$, are defined in Section~\ref{metrics}. }
  \label{fig:evaluation_metrics}
\end{figure}

Second, to evaluate the distribution of the generated dataset, 
the relationship between the J2W-based model and the OWA-based model is identified and we confirm whether JABBERWOCK can generate samples close to those in the real world. 
Namely, we focus on the difference between the cosine similarities of the generated samples. 
While cosine similarity is a typical metric to evaluate the similarity between vectors output by a model, 
it is considered that the above two models can vectorize inputs equivalently between them as long as the difference between cosine similarities of the vectors is close to zero. 
That is, even if each model outputs vectors in different directions, it can still identify whether the relationship between these vectors is similar, as shown in Fig.~\ref{fig:evaluation_metrics}. 
We also assume that both J2W-based and OWA-based models are unlabeled in this experiment to evaluate the difference between cosine similarities.

Third, we confirm the accuracy of datasets generated by JABBERWOCK by detecting malicious websites.
This confirms that WebAssebly based malicious website detection through JABBERWOCK is effective.

\subsection{Experimental Setting}
We describe data collection and metrics in the experimental setting below. 
We implemented JABERWOCK with Python version 3.7.10 on the AWS EC2 Ubuntu t3a.xlarge instance. 

\subsubsection{
Data Collection}
For the first experiment, we obtained an URL list from PublicWWW\footnote{\url{https://publicwww.com}}.
It contains 10,000 samples of JavaScript for J2W and 168 samples of WebAssembly for OWA.
We remark that the URL list is not labeled in this experiment.
This is because there are few (less than 10\%) malicious WebAssembly samples among the genuine samples. 
We measured the processing time (Section~\ref{sec:process_time}), and compared J2W and OWA (Section~\ref{sec:diff_cosinesimilarity}).

For the second experiment, i.e. malicious website detection, we prepared an input to JABBERWOCK as follows.
Benign websites were obtained from Tranco~\cite{tranco}, and malicious websites were obtained from existing malicious website lists such as URLhaus\footnote{\url{https://urlhaus.abuse.ch}}, Malware Domain List\footnote{\url{https://www.malwaredomainlist.com/mdl.php}}, CyberCrime Tracker\footnote{\url{https://cybercrime-tracker.net}}, and PhishTank\footnote{\url{https://phishtank.org}}.
These data were classified into 80\% of training data and 20\% of test data, respectively. 
We gave JABBERWOCK 1783 benign websites and 1783 malicious websites to obtain a labeled dataset. 
We trained machine-learning models based on the labeled dataset. 
Finally, we confirmed that the models can be used for malicious website detection by giving the test data to the trained models. 
For more details, see Section~\ref{sec:disucssion_accuracy}.

\subsubsection{Evaluation Metrics}

As evaluation metrics, we focus on the dataset generation by JABBERWOCK 
and the detection accuracy for the malicious website detection are described below.

\paragraph{Performance of JABBERWOCK
}
\label{metrics}

We evaluate the performance of dataset generation by JABBERWOCK with respect to the processing time and the distribution of the generated dataset. 
We describe the latter in detail below. 

For the evaluation of the distribution of the generated dataset, we introduce a new metrics that extends the existing evaluation metrics, the cosine similarity.
We define some variables to describe the metrics.
Let $\mathrm{J}_1, \cdots, \mathrm{J}_n$ (resp., $\mathrm{O}_1, \cdots, \mathrm{O}_m$) be WATs that are obtained from test data of J2W (resp., OWA).
In what follows, we do not mention $\mathrm{O}_i$'s, because they are treated in the same manner as $\mathrm{J}_i$'s.
For each $i = 1, \cdots, n$, the vector obtained by processing $\mathrm{J}_i$ by J2W-based model is called $\mathrm{J}_i$\_J2W.
We define $\mathrm{J}_i$\_OWA in the same way. 

To evaluate the dataset generated by JABBERWOCK, we first compute cosine similarities for each combination of $\mathrm{J}_i$ and $\mathrm{J}_j$.
We denote the cosine similarity between $\mathrm{J}_i$ and $\mathrm{J}_j$ by J2W-based model (resp., by OWA-based model) by $\mathrm{J}_i \mathrm{J}_j$\_J2W (resp., $\mathrm{J}_i \mathrm{J}_j$\_OWA). 
Recall that we aim to evaluate similarities between vectors of JABBERWOCK and the actual WebAssembly samples.
We then evaluate the difference between cosine similarities as $|(\mathrm{J}_i \mathrm{J}_j\mathrm{\_J2W})-(\mathrm{J}_i \mathrm{J}_j\mathrm{\_OWA})|$.
We compute the differences for each combination of $i \neq j$, and evaluate them through a box-and-whisker plot. 
Intuitively, the smaller the difference is, the better the result is.

\paragraph{Accuracy of Malicious Website Detection}

We explain the evaluation metrics in malicious website detection.
We first describe the four terms, i.e., true positive (TP), true negative (TN), false positive (FP), and false negative (FN), to evaluate the detection performance.
TP is the number of malicious websites that are correctly detected as malicious. 
TN is the number of benign websites that are correctly detected as benign. 
FP is the number of benign websites that are wrongly detected as malicious, and FN is the number of malicious websites that are wrongly detected as benign.
We then define the following evaluation metrics based on the above four terms.
\begin{itemize}
    \item Accuracy: It is the ratio of correctly detected websites to the total number of websites.
    
    \item Precision: It is the ratio of the number of correctly detected websites as malicious to the total number of detected websites as malicious.
    
    \item Recall: It is the ratio of the number of correctly detected websites as malicious to the total number of malicious websites. 
    
    \item F1 score: It is the harmonic mean of the precision and the recall.
    
\end{itemize}

\subsubsection{Hyperparameters}


When vectorization is performed by Doc2Vec, the hyperparameters are set as follows.
The vector size of Doc2Vec is set to be $2,4,6,8,10,20,40,60,80$, and $100$.
We choose either PVDM or PVDBOW as a model of Dco2Vec. 
In the default setting, the vector size is 100, the number of epochs is 10 and the model is PVDM.
For malicious website detection, we also used support vector machines with radial basis functions, polynomial functions, and linear functions as kernel (SVM\_rbf, SVM\_poly, SVM\_linear), Random Forest (RF), Naive Bayes (NB), XGBoost (xgboost), and LightGBM (lightgbm) were used.

\subsection{Results} \label{subsec:results}

Experimental results based on the evaluation metrics described in the previous section are described below.

\subsubsection{Performance of JABBERWOCK} \label{sec:eval_jabberwock_process}

The processing time and the distribution evaluation of the datasets generated by JABBERWOCK are described below.

\paragraph{Processing Time} \label{sec:process_time}

\begin{figure}[tb]
 \centering
  \includegraphics[width=\linewidth]{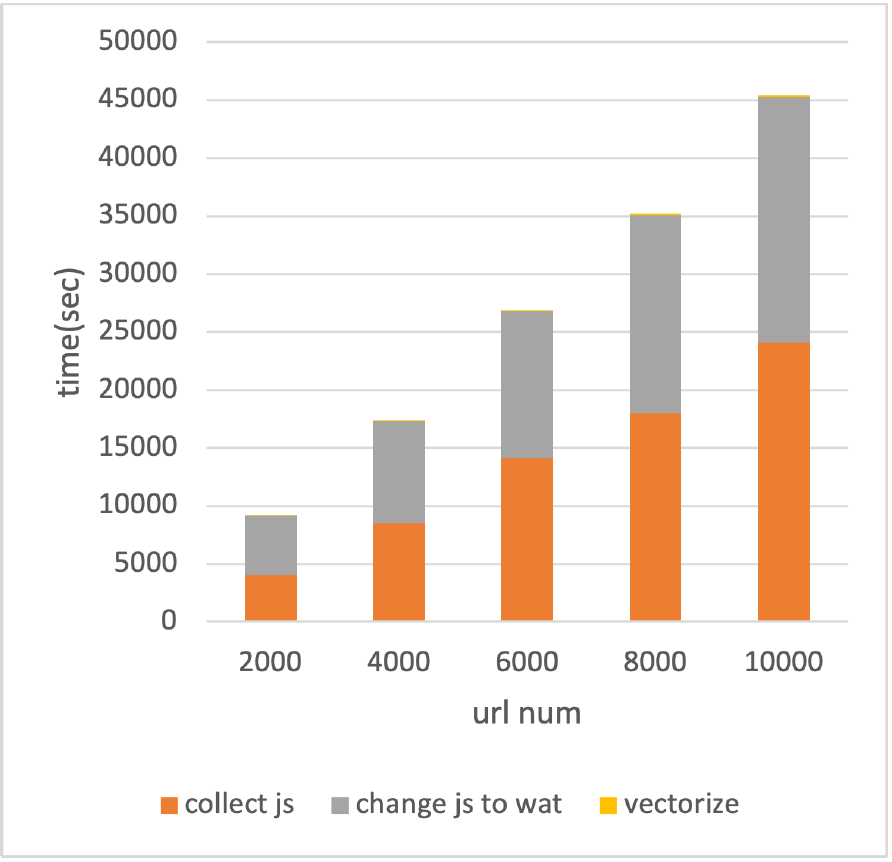}
  \caption{The processing time of JABBERWOCK: Each bar contains three processes of JABBERWOCK. 
  The vectorization process is quite short compared to the other two processes.}
  \label{time:url}
\end{figure}

Fig.~\ref{time:url} demonstrates the processing time of JABBERWOCK for each number of URLs of malicious websites. 
Increasing the number of URLs to be converted by 2000, experimenting from 2000 to 10000.
Compared with the vectorization process (\textsf{vectorize}), the whole processing time of JABBERWOCK is occupied by the collection of JavaScripts (\textsf{collect js}) and translating to WAT  (\textsf{change js to wat}). 
These times increase in proportion to the number of URLs.
This show that JABBERWOCK can construct a dataset in 4.5 seconds per sample.
This process is once for all task, so this time is not long.

\begin{figure}[tb]
 \centering
  \includegraphics[width=1\linewidth]{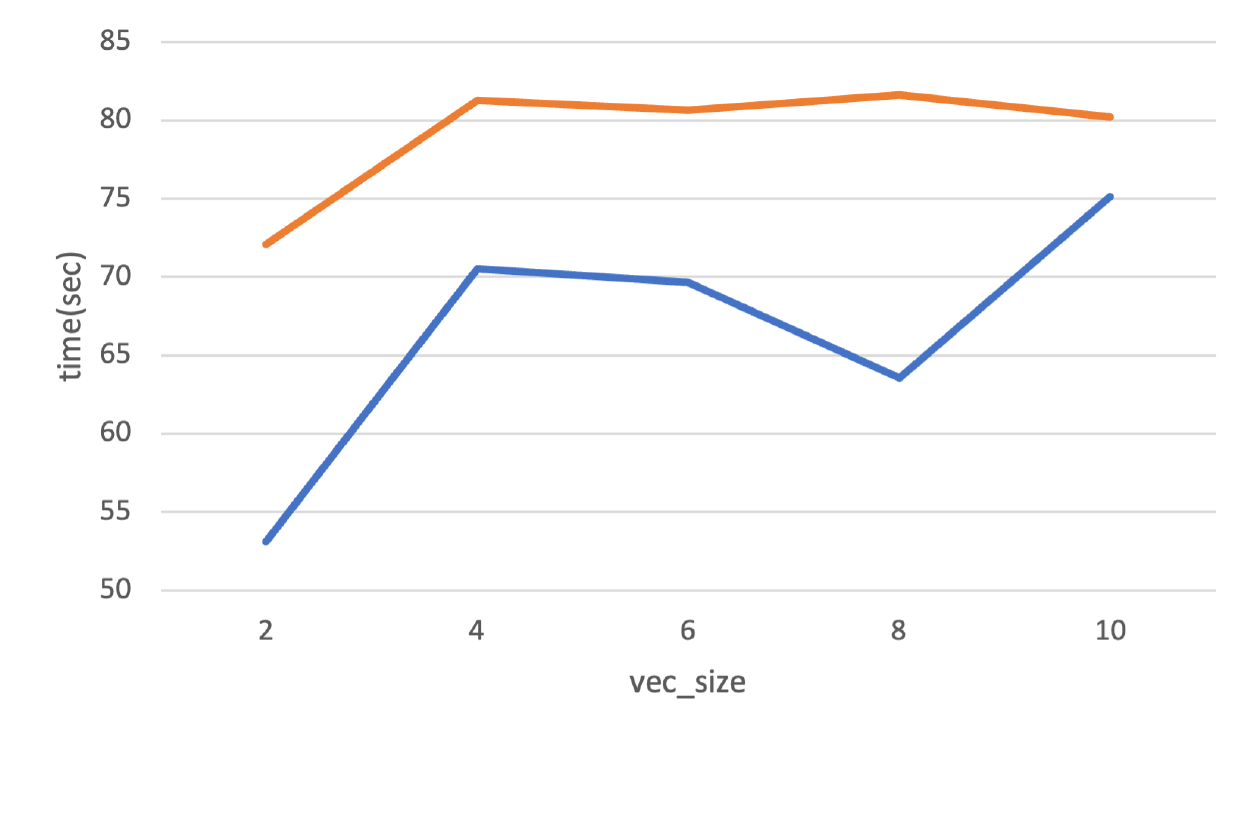}
  \caption{The processing time for the vectorization process when the vector size is small: 
  The orange line represents PVDM and the blue line represents PVDBOW as Doc2Vec, respectively.}
  \label{time:vec2}
\end{figure}

\begin{figure}[tb]
 \centering
  \includegraphics[width=1\linewidth]{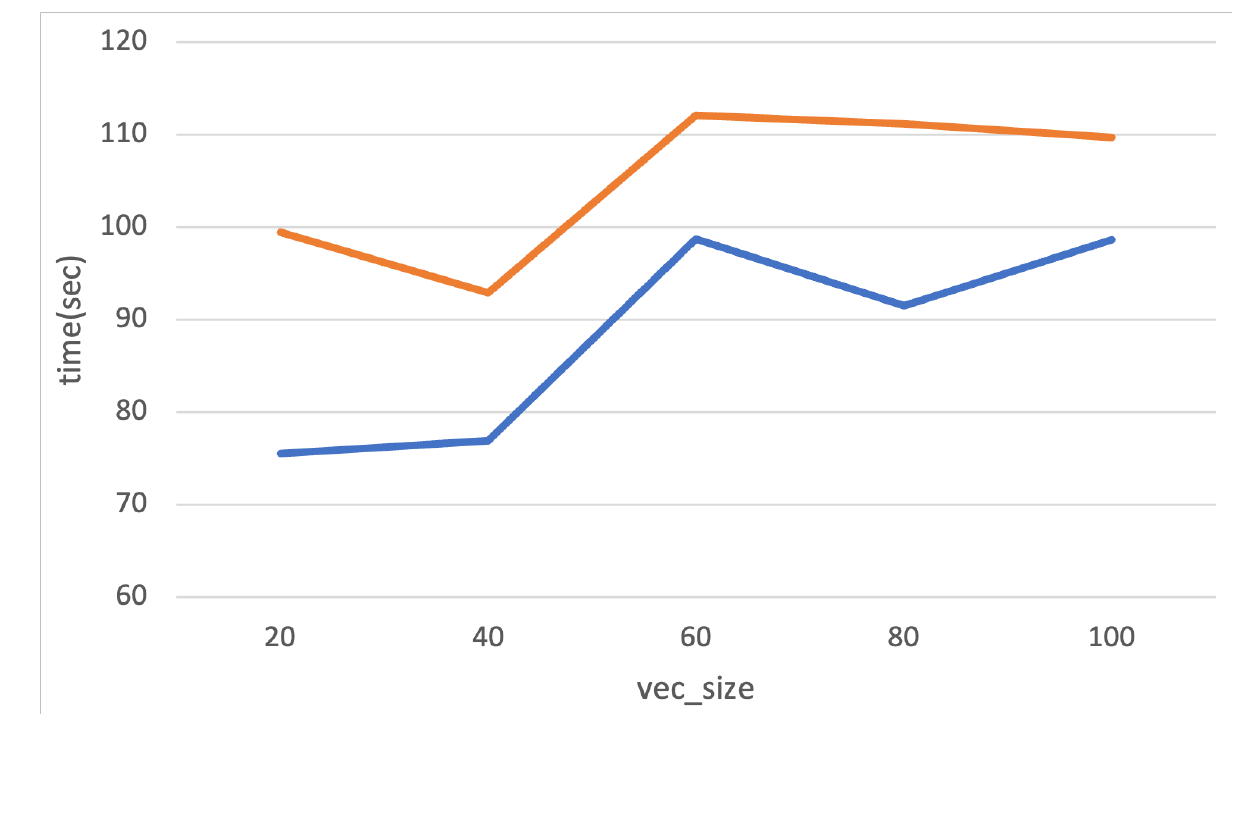}
  \caption{The processing time for the vectorization process when the vector size is large: 
  The orange line represents PVDM and the blue line represents PVDBOW as Doc2Vec, respectively.}
  \label{time:vec20}
\end{figure}

In processing time, vectorization can vary with hyperparameters, so we focus on the processing time of vectorization.
Specifically, we investigated the case when the vector size is small in detail, as it might affect the processing time. 
Thus, we demonstrate two figures; Fig.~\ref{time:vec2} illustrates the processing time when the vector size is $2, 4, 6, 8$ and $10$, and Fig.~\ref{time:vec20} illustrates when $20, 40, 60, 80$ and $100$.
Fig.~\ref{time:vec2} shows the processing time for the vectorization.
The processing time tends to increase gentry in proportion to the vector size. 
We remark that this tendency continues even when the vector size is varied up to 100 from Fig.~\ref{time:vec20}.
Furthermore, processing in PVDBOW takes a longer time than processing in PVDM. 

\paragraph{
Distribution of Generated Dataset
} \label{sec:diff_cosinesimilarity}



\begin{figure}[t]
 \centering
  \includegraphics[width=1\linewidth]{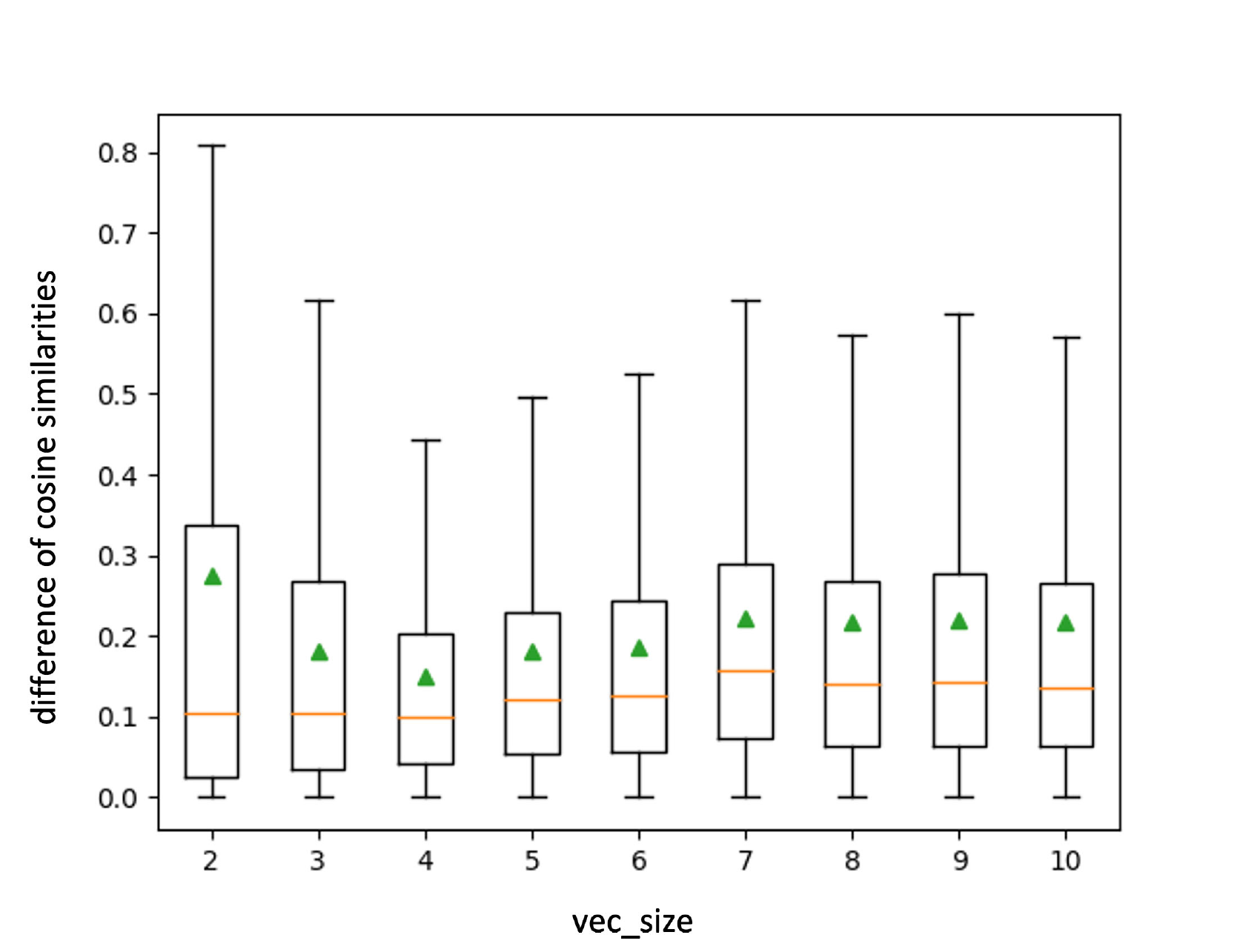}
  \caption{Difference between cosine similarities when the vector size is small.}
  \label{fig:vec2}
\end{figure}

\begin{figure}[t]
 \centering
  \includegraphics[width=1\linewidth]{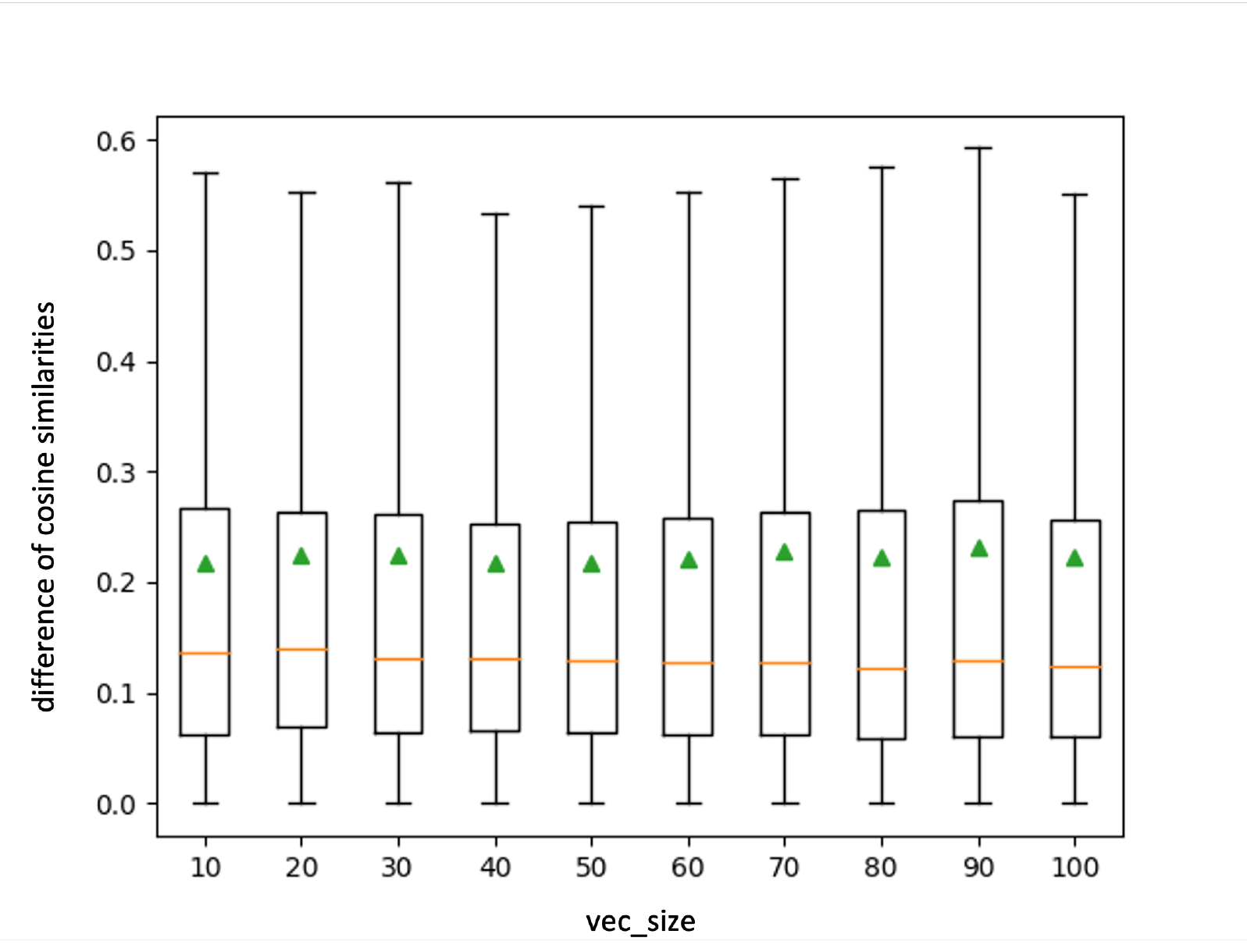}
  \caption{Difference between cosine similarities when the vector size is large.}
  \label{fig:vec10}
\end{figure}

Fig.~\ref{fig:vec2} shows the difference in cosine similarities between the J2W-based and OWA-based models for evaluation of the distribution of the generated dataset. 
Recall that the default model is PVDM. 
Similar to the processing time evaluation, we investigated the case when the vector size is small in detail. 
When the vector size is 2, the variance becomes large, and hence the generated samples are unstable compared to samples in the real world. 
However, as the size of the vectors increases, the differences in cosine similarities become stably smaller.
Intuitively, the vectors generated by JABBERWOCK tend to be more stably smaller than those extracted from WebAssembly in the real world.
The above phenomenon is also seen even when the vector size is larger than 10 from Fig.~\ref{fig:vec10}. 

\begin{figure}[t]
 \centering
  \includegraphics[width=1\linewidth]{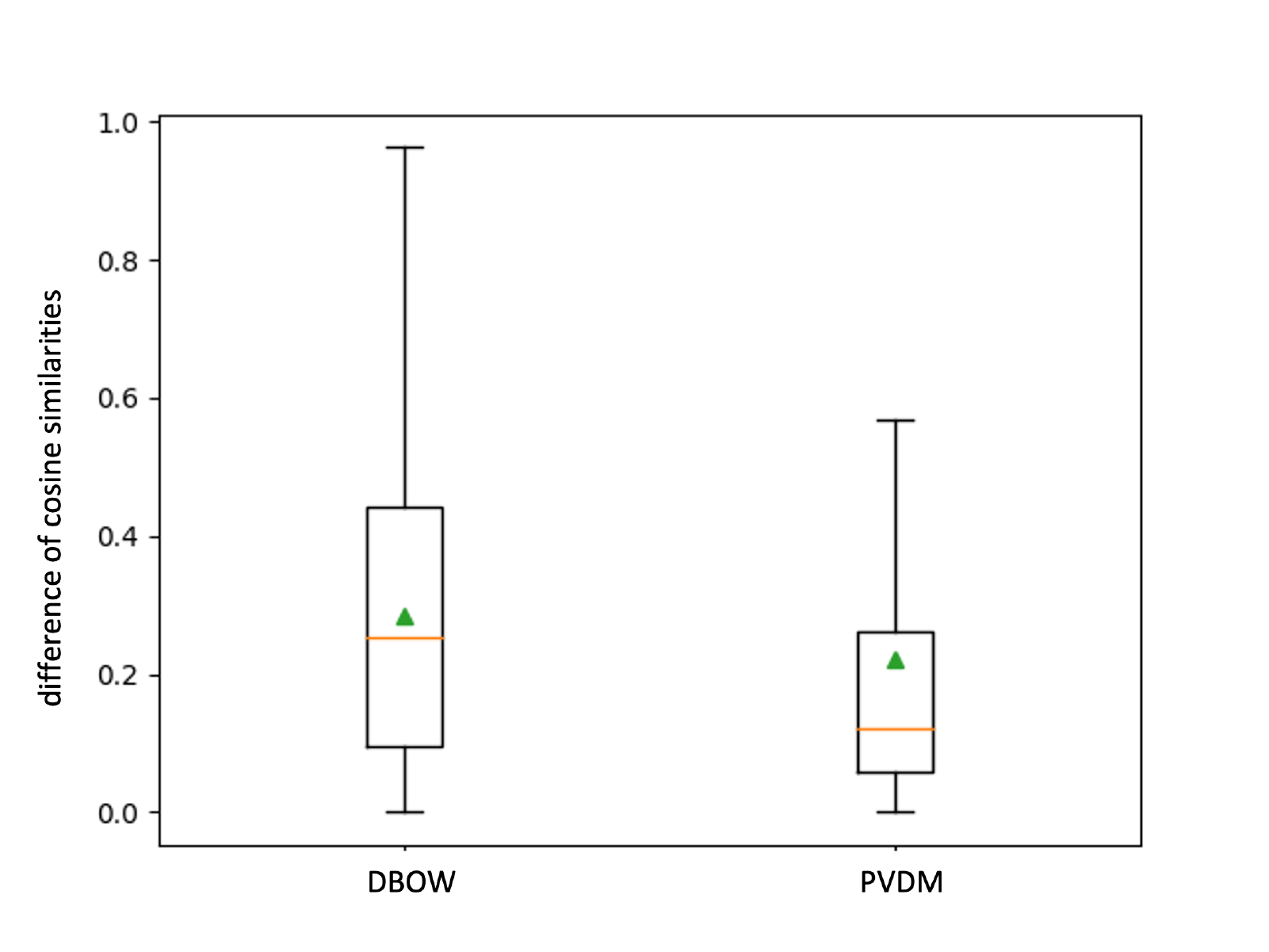}
  \caption{Difference between cosine similarities when the model is PVDM and PVDBOW, respectively.}
  \label{fig:dm}
\end{figure}

Fig.~\ref{fig:dm} demonstrates the difference between cosine similarities when the model is PVDBOW and PVDM, respectively. 
Compared with the case of PVDBOW, the variance is smaller in the case of PVDM. 
It indicates that the generated samples by JABBERWOCK are more similar to WebAssembly codes in the real world for PVDM. 
Hence, JABBERWOCK can potentially generate samples similar to WebAssembly codes in the real world. 


\subsubsection{Accuracy of Malicious Website Detection} \label{sec:disucssion_accuracy}

\begin{figure*}[tb]
   \centering
   \begin{tabular}{cc}
       \begin{minipage}{0.5\hsize}
           \centering
          \includegraphics[width=\linewidth]{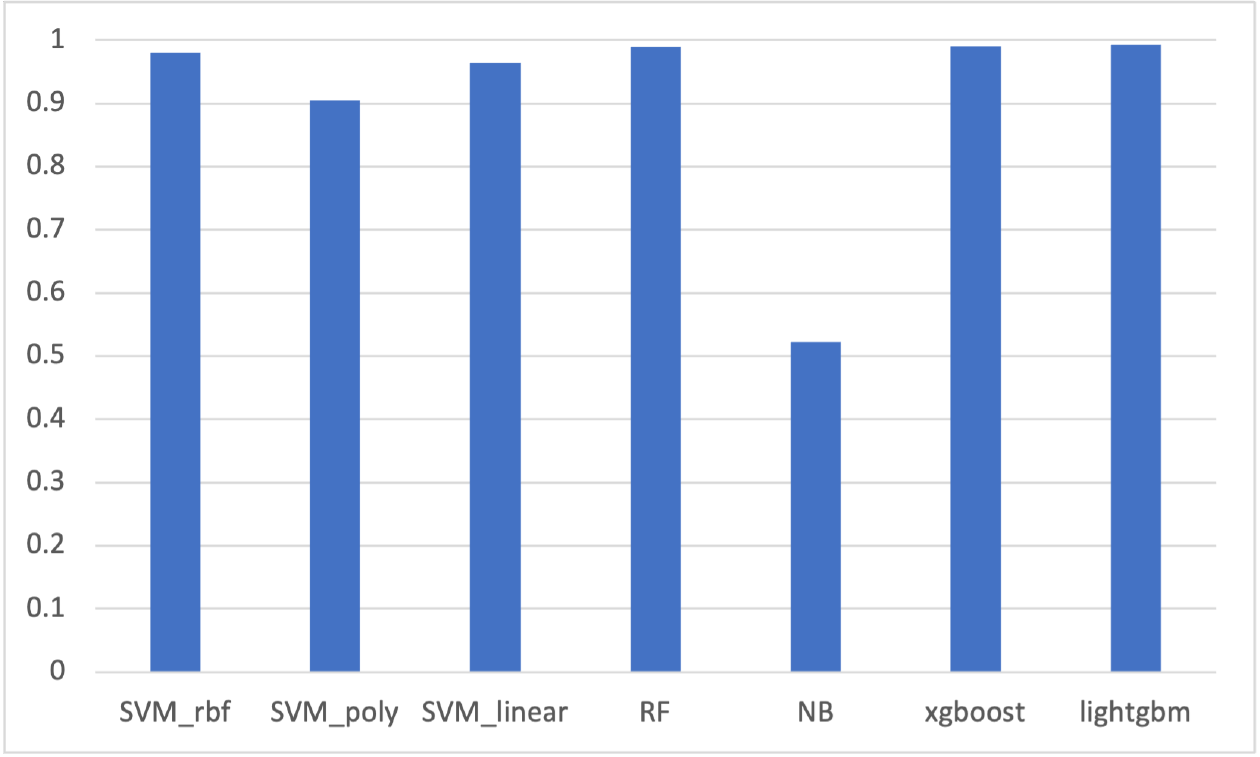}
          \subcaption{F1 score}
          \label{f1}
       \end{minipage}&

       \begin{minipage}{0.5\hsize}
           \centering
           \includegraphics[ width=1\linewidth]{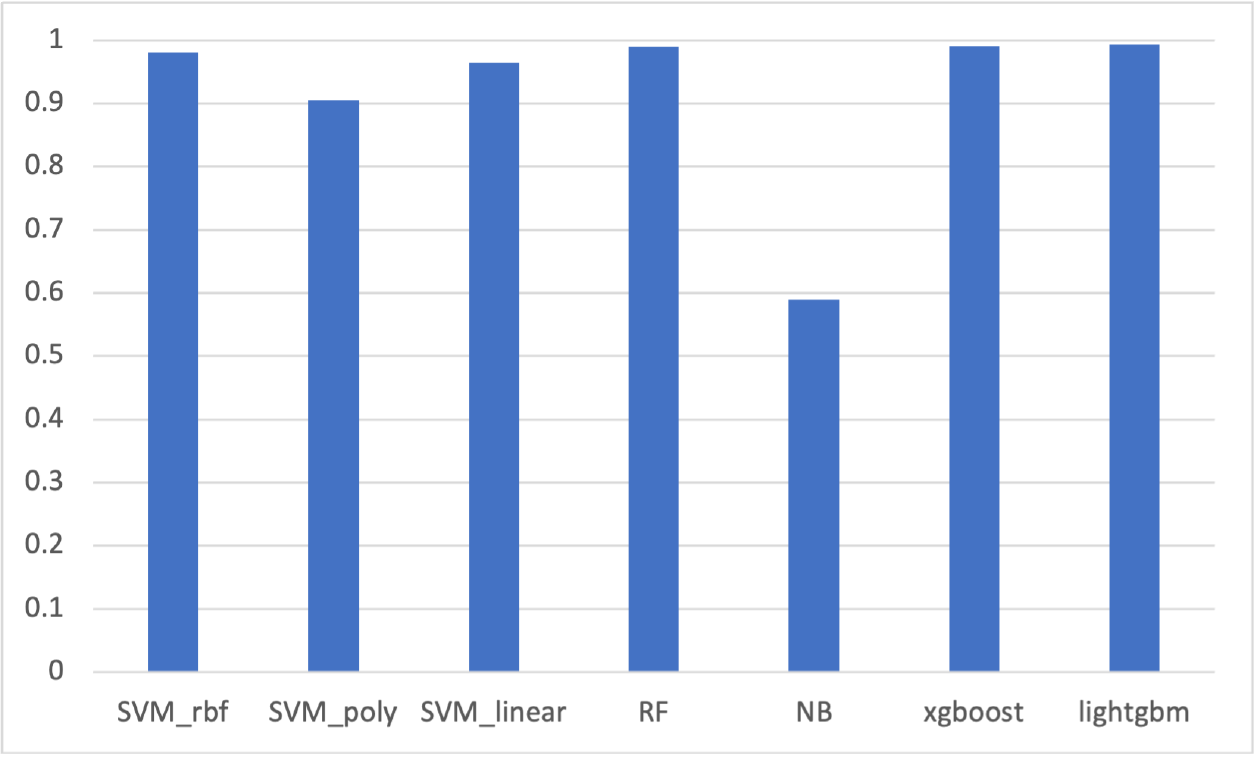}
           \subcaption{Accuracy}
            \label{accuracy}
       \end{minipage}\\

       \begin{minipage}{0.5\hsize}
           \centering
          \includegraphics[width=\linewidth]{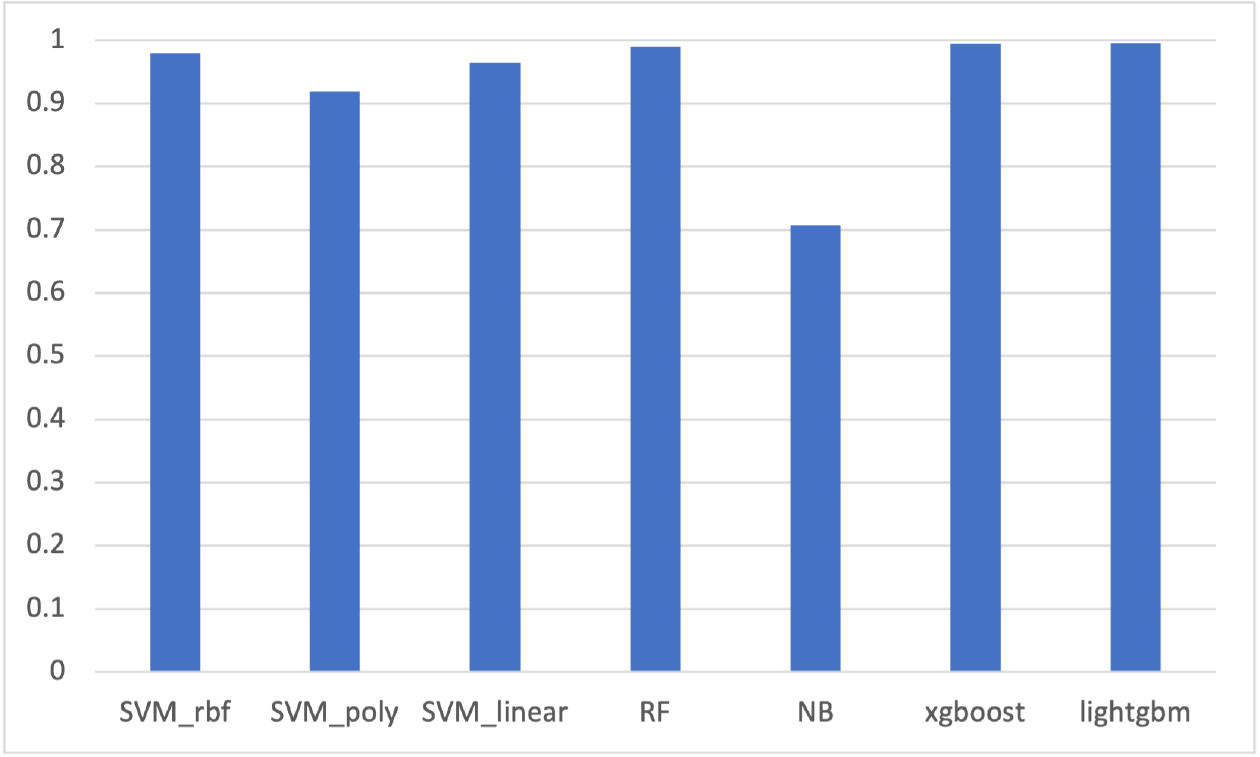}
          \subcaption{Precision}
          \label{orecision}
       \end{minipage}&

       \begin{minipage}{0.5\hsize}
           \centering
           \includegraphics[ width=1\linewidth]{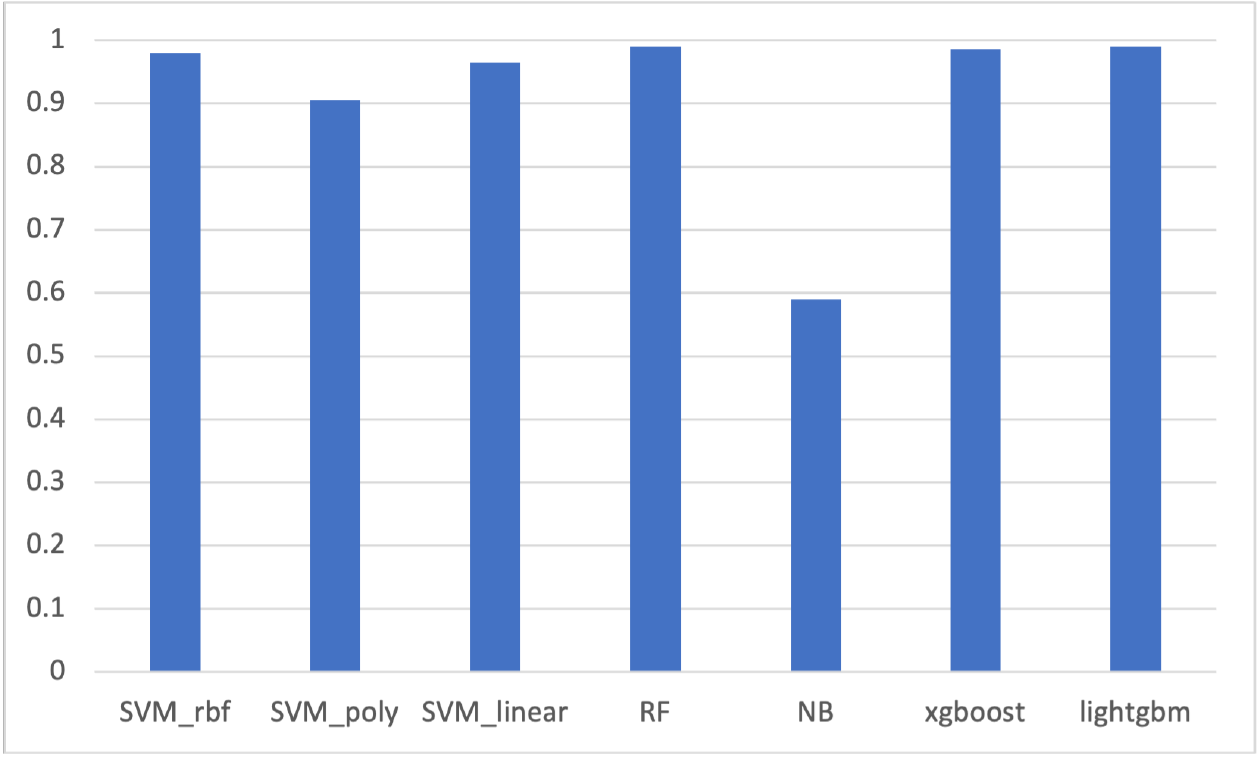}
           \subcaption{Recall}
            \label{recall}
       \end{minipage}
       
   \end{tabular}
   \caption{Detection accuracy by model}
   \label{fig:accuray_evaluation}

\end{figure*}

Fig.~\ref{fig:accuray_evaluation} shows the results of the evaluation of the accuracy of detecting malicious websites.
In the following results, initial values are used as hyperparameters of Doc2Vec.
The accuracy did not change much depending on the parameters.

First, Fig.~\ref{f1} shows that support vector machines with radial basis functions, Random Forest, XGBoost, and LightGBM have the highest F1 score, reaching 99.6\%.
Similar trends are observed in the accuracy, precision, and recall from Fig.~\ref{accuracy}$\sim$\ref{recall}.
On the other hand, for the support vector machine with polynomial functions and Naive Bayes, all the values are significantly lower than those of the other models.
The reasons for the accuracy of each graph are discussed in the next section, but the overall trend of each graph indicates that malicious website detection is successful.

\section{Discussion} 

Based on the results of the previous section, we discuss the reason for the high detection accuracy of malicious website detection with the dataset generated by JABBERWOCK.
First, we analyze the reason for detection accuracy described in the previous section.
Next, we also discuss integration with an existing tool~\cite{MADMAX} and 
the impact of the hyperparameters of JABBERWOCK for malicious website detection. 
We finally discuss the cybersecurity ethics and limitations of this paper.

\subsection{Analysis of Results}

In this section, we discuss the reasons for the overall high detection accuracy of malicious websites described in Section~\ref{sec:disucssion_accuracy}, and the reasons for the significantly low Naive Bayes accuracy.

\begin{figure*}[tb]
   \centering
   \begin{tabular}{cc}
       \begin{minipage}{0.5\hsize}
           \centering
          \includegraphics[width=\linewidth]{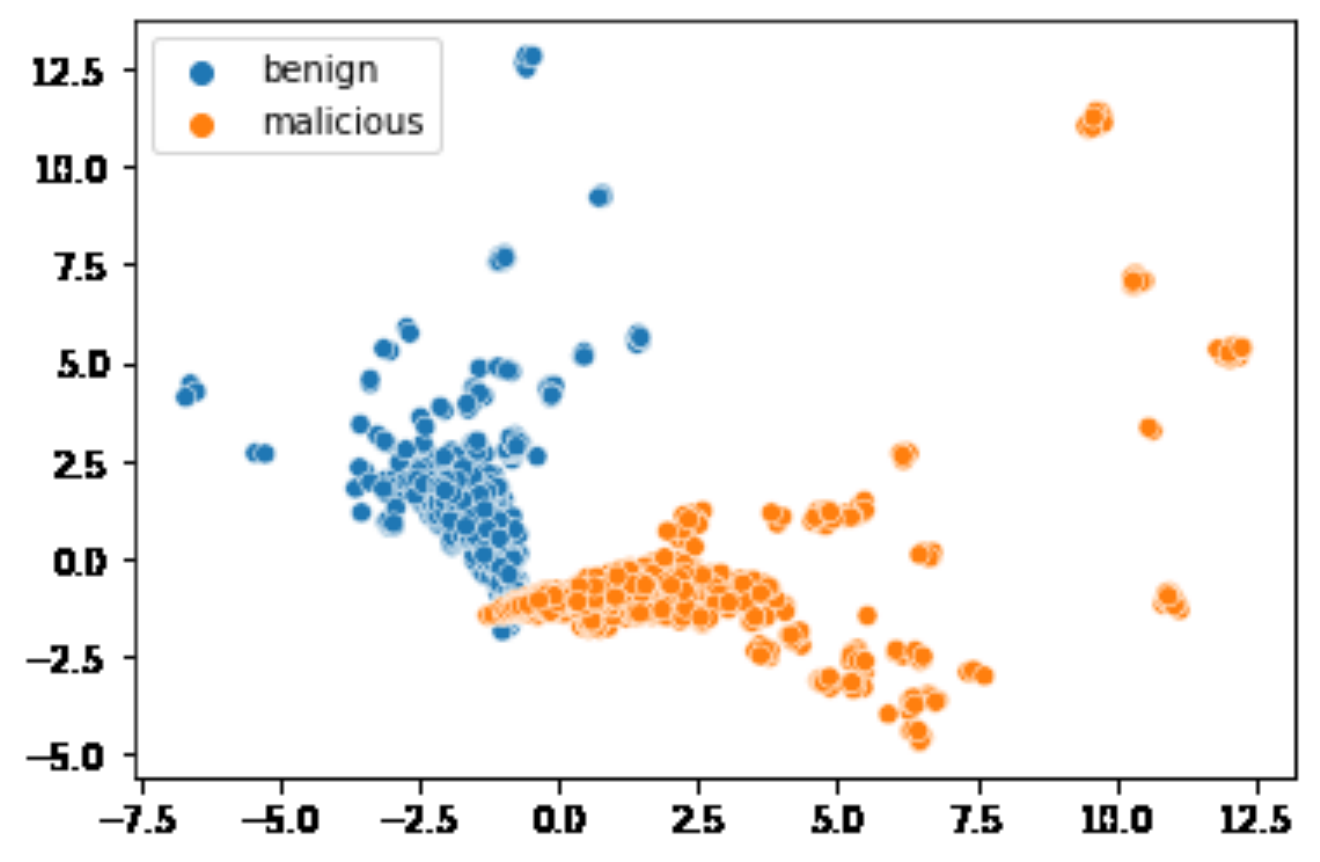}
          \subcaption{Samples generated by JABBERWOCK}
          \label{scatter}
       \end{minipage}&

       \begin{minipage}{0.5\hsize}
           \centering
           \includegraphics[ width=1\linewidth]{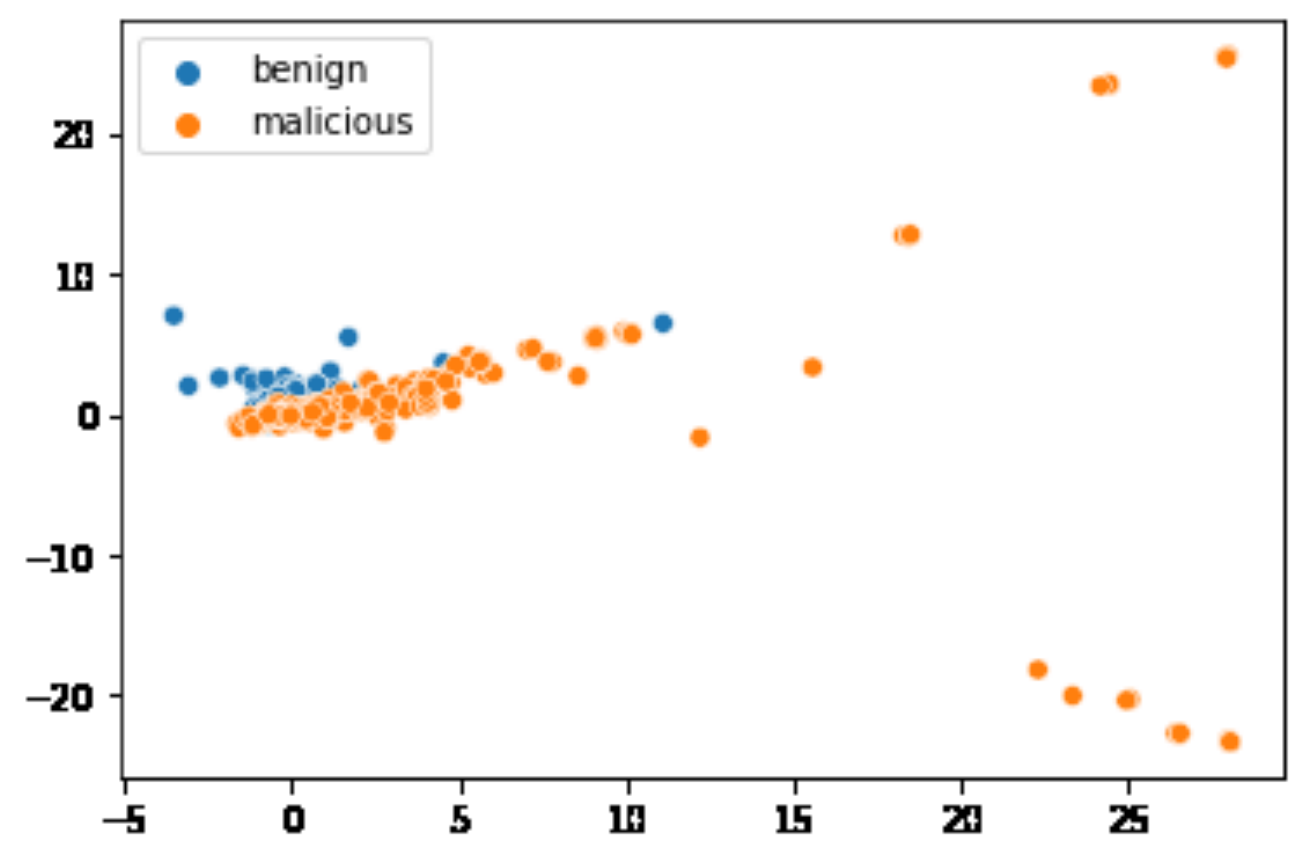}
           \subcaption{Original JavaScript code}
            \label{js_scatter}
       \end{minipage}\\

   \end{tabular}
   \caption{Distribution of samples: We measure distributions of samples, where the principal component analysis is utilized to compress dimensions of the samples. 
   The left figure represents the samples generated by JABBERWOCK and the right figure represents their original JavaScript codes.}

\end{figure*}

First, we discuss the reasons for the high accuracy of the support vector machine with radial basis functions, Random Forest, XGBoost, and LightGBM.
In particular, we analyze the distribution of vectors obtained by JABBERWOCK.
As the beginning of the analysis, we conduct principal component analysis on the vectors, we compress the vectors, which initially have 100 dimensions, to two dimensions using the \texttt{scikit-learn} library.
The results are visualized by plotting them on a figure with the \texttt{seaborn} library.
Fig.~\ref{scatter} is the result of scatterplot.
This figure shows the distribution of the data, and we would like readers to pay attention to the plotted state of each sample rather than the numerical values on the axes.


Fig.~\ref{scatter} shows that the distributions of benign and malicious websites are far apart.
Therefore, we can consider that the accuracy of detecting malicious websites is significantly improved.

A similar analysis was performed for the original JavaScript, and an interesting phenomenon was confirmed.
Fig.~\ref{js_scatter} shows the distribution of the original JavaScript.
Comparing Fig.~\ref{scatter} and Fig.~\ref{js_scatter}, we can see that the vector distribution obtained by JABBERWOCK shows a larger discrepancy in the distance between benign websites and malicious websites than the vector distribution of the original JavaScript.
This suggests that the rend of benign and malicious features has become clear through the conversion to WebAssembly by JABBERWOCK.
We think that the reason for the large discrepancy in distribution is that the Wobfuscator conversion splits the code into two parts: the part that can be converted to WebAssembly and the part that remains in JavaScript.
We believe that when the split is made, the code that is the key dynamic part of the web page will be included in WebAssembly.


The reason for the low detection accuracy of Naive Bayes is considered to be due to the characteristics of Naive Bayes.
While Naive Bayes can reduce the number of parameters, it cannot reflect the correlation between features in the results, and it assumes that each feature is independent~\cite{generativedeeplearning}.
For this reason, especially when the size of vectors is increased in Doc2Vec, it is considered that learning was not properly performed.
For use cases of JABBERWOCK, it is better to use XGBoost or LightGBM.

\subsection{Integration with Existing Tools} 


To confirm the effectiveness of the WebAssembly dataset generated by JABBERWOCK against actual malicious website detection techniques, we compare and evaluate JABBERWOCK with the latest malicious website detection technique, MADONNNA~\cite{janaka2023madonna}, which does not learn WebAssembly codes.
First, 1338 benign websites and 1338 malicious websites were obtained for each of the features described in the literature~\cite{MADMAX}, using the data collection method used in  Section~\ref{sec:purpose_of_experiments}.
Here, websites for which a WebAssembly sample was not generated by JABBERWOCK were excluded as missing values.
The benign cases were selected from the top of the list.

We measured the accuracy of malicious website detection using WebAssembly generated by JABBERWOCK in combination with features from the literature~\cite{MADMAX} and obtained the results shown in Table~\ref{tab:comapre_with_madonna}.
The table shows that incorporating WebAssembly improved the F1 score by more than 9 points.
This is expected to improve the accuracy when JABBERWOCK is combined with existing malicious website detection techniques.

\begin{table}[htb]
    \centering
    \caption{Comparison with existing work for F1-score: 
    We simply refer to the scores shown in~\cite{janaka2023madonna}. 
    }
    \begin{tabular}{c|c|c} \hline
        Models & Our results & MADONNA~\cite{janaka2023madonna} \\
        \hline
        RF & 0.99 & 0.87 \\
        XGBoost & 0.99 & 0.90 \\
        LightGBM & 0.99 & 0.89 \\
        \hline
    \end{tabular}
    \label{tab:comapre_with_madonna}
\end{table}


    


\subsection{Impact of Hyperparameters}

We discuss how the JABBERWOCK hyperparameters affect the generated vectors from the results of  Section ~\ref{sec:process_time} and  Section~\ref{sec:diff_cosinesimilarity}.
From each result, we can make the following two statements.
First, the processing time increases in proportion to the size of the vector, while the distribution of the differences among the cosine similarities is almost the same when the size of the vector is larger than 10, as mentioned above.
Therefore, the appropriate size of the vector is 10.
Next, when PVDM is selected for Doc2Vec, the processing time increases due to the training of PVDM, but as mentioned above, the resulting differences among the cosine similarities tend to be stable with smaller variance.
For this reason, it is preferable to use PVDM rather than PVDBOW.

\subsection{Cybersecurity Ethics} 

We discuss cybersecurity ethics in this paper. 
When we collected actual JavaScript code, we utilized only popular ranking websites which are publicly available. 
The collection was performed in our local environment. 
Then, we affect neither the underlying services of these websites nor their contents. 
In doing so, the contents were collected by the standard access to websites, i.e., \texttt{HTTP POST} and \texttt{GET} methods. 
Namely, we were able to collect data that are identical to the ones obtained by their standard usage from the websites. 
Consequently, we can ensure that no personal data is collected and the user's privacy is guaranteed. 

\subsection{Limitations} 

JABBERWOCK has three limitations, i.e., missing values, an application for real-time detection, and labeling samples.

First, we did not care about the missing values. 
When no JavaScript codes are found or translated into WebAssembly codes, a dataset will no longer be generated. 
In the current experiments, we ignore the missing values and still have the possibility whereby the performance of JABBERWOCK can be improved by dealing with them in a better fashion. 
Second, an application with JABBERWOCK for real-time detection seems to be difficult. 
As shown in Section~\ref{sec:eval_jabberwock_process}, 
we need to take 4.5~seconds to convert JavaScript to WebAssembly codes. 
It is longer than a real-time detection application such as MADMAX~\cite{MADMAX}. 
Third, labeling samples straightforwardly follow the setting of the existing work~\cite{MADMAX}. 
We also need to confirm the validity of this setting for labeling samples. 

Further studies, which take the above limitations into account, will need to be undertaken.

\section{Conclusion} 

In this paper, we designed JABBERWOCK, a tool for generating a WebAssembly dataset, and then conducted experiments for malicious website detection with the dataset generated by JABBERWOCK. 
The main procedure of JABBERWOCK is to gather JavaScript codes from malicious websites and then convert them into WebAssembly. 
We then evaluated JABBERWOCK with respect to processing time and the difference between models trained with the generated dataset. 
We showed that JABBERWOCK could generate any number of samples with 4.5 seconds per sample. 
We also demonstrated that samples generated by JABBERWOCK are close to WebAssembly samples in the real world. 
Furthermore, we evaluated the performance of malicious website detection with the dataset generated by JABBERWOCK and then showed that malicious websites could be detected with 99\% F1-score. 
When we analyzed the reason for such a high score, we found the fact that JABBERWOCK makes a gap between benign and malicious samples larger than that of the original JavaScript codes. 
We also evaluated the integration with the existing tool~\cite{janaka2023madonna} and then confirmed that the F1-score can be improved. 
We thus confirmed that JABBERWOCK is effective, i.e., malicious website detection based on WebAssembly is possible. 
In the future, we plan to improve the processing time for the conversion of JavaScript into WebAssembly in order to provide a browser-based real-time application for malicious website detection, such as MADMAX~\cite{MADMAX}. 



\textbf{Reproducibility:} 
The code of JABBERWOCK is available via GitHub (\url{https://github.com/c-chocolate/Jabberwock}) except for Wobfuscator~\cite{Wobfuscator}. 
(Obtain the code of Wobfuscator from Dr. Alan Romano, the corresponding author of Wobfuscator.)

\textbf{Acknowledgement:} 
JST CREST, the grant number JPMJCR21M5, supports a part of this work. 
We would also appreciate Dr. Alan Romano for providing his code of Wobfuscator. 






\end{document}